\documentclass[12pt]{article}
\usepackage{graphicx,amsmath, amsthm, amssymb, natbib, color,graphicx, hyperref, amsfonts, sectsty, caption, subcaption,
mathrsfs,bbm,float,authblk}
\usepackage{psfrag,epsf}
 \usepackage[dvipsnames]{xcolor}
\usepackage{enumitem,tcolorbox,soul}
\usepackage{xr}
\def\nhalf{n^{\frac{1}{2}}}

\def\spacingset#1{\renewcommand{\baselinestretch}%
{#1}\small\normalsize} \spacingset{1}

\usepackage{marquis_general}
\usepackage{double_surrogate_notation}

\usepackage[dvipsnames]{xcolor}

\usepackage{tikz}
\usetikzlibrary{arrows}
\usetikzlibrary{shapes.geometric}
\usetikzlibrary{fit}

\usepackage[margin=1in]{geometry}
\allowdisplaybreaks

\definecolor{jcolor}{RGB}{041,122,000}
\definecolor{darkblue}{RGB}{000,000,150}
\definecolor{darkred}{RGB}{100,000,000}
\definecolor{purple}{RGB}{200,000,200}

\begin{document}


\author[1]{Jue Hou}
\author[1]{Rajarshi Mukherjee}
\author[1,2]{Tianxi Cai}
\affil[1]{Department of Biostatistics, Harvard T. H. Chan School of Public Health}
\affil[2]{Department of Biomedical Informatics, Harvard Medical School}

\title{Efficient and Robust Semi-supervised Estimation of ATE
 with Partially Annotated Treatment and Response}

\date{}

\maketitle

\bigskip
\begin{abstract}
A notable challenge of leveraging Electronic Health Records (EHR) for treatment effect assessment is the lack of precise information on important clinical variables, including the treatment received and the response. Both treatment information and response often cannot be accurately captured by readily available EHR features and require labor intensive manual chart review to precisely annotate, which limits the number of available gold standard labels on these key variables. We consider average treatment effect (ATE) estimation under such a semi-supervised setting with a large number of unlabeled samples containing both confounders and imperfect EHR features for treatment and response. We derive the efficient influence function for ATE and use it to construct a semi-supervised multiple machine learning (SMMAL) estimator. We showcase that our SMMAL estimator is semi-parametric efficient with B-spline regression under low-dimensional smooth models. We develop the adaptive sparsity/model doubly robust estimation under high-dimensional logistic propensity score and outcome regression models. Results from simulation studies support the validity of our SMMAL method and its superiority over supervised  benchmarks.
\end{abstract}
\noindent%

\section{Introduction}

 The 21st Century Cures Act and the Prescription Drug User Fee Act VII have shone spotlight on the use of real-world evidence, generated from real-world data, to support regulatory-decision making on drug effectiveness. Large scale electronic health records (EHR) data are being increasingly used for creating the real-world evidence for treatment effectiveness or efficacy \citep{FranklinEtal21history}. Using EHR data to infer treatment effect, however, is highly challenging. In addition to the observational nature, another notable challenge in leveraging EHR for treatment effect assessment lies in the curation of key clinical variables, including the treatment being investigated and the outcome of interest.
Response variables such as disease progression may not be well represented by readily available EHR features \citep{BartlettEtal19}. Treatment information can be partially captured but not always accurately reflected by medication prescription codes. New therapies may not be well coded in the early adoption stage and treatment initiation may be later than prescription date due to external factors such as insurance approval delay.

Although it is possible to improve treatment or response definition by combining multiple EHR features through rule based or machine learning algorithms, these EHR derived features are at best good surrogates for approximating the true treatment or response information at patient level. Directly using these surrogates for treatment and outcome which would potentially induce bias in the subsequent analysis \citep{BJEtal20}. On the other hand, precisely curating treatment and response variables via manual chart review by domain expert is resource intensive, leading to limited sample size for gold standard labels on these key variables. It is thus of great interest to develop robust semi-supervised learning (SSL) procedures to derive unbiased and efficient inference about the average treatment effect (ATE), leveraging both the small number of gold standard labels and the large sized unlabeled data on the surrogates and confounders.

Additional challenges arise from the high dimensionality of potential confounders.
Unlike traditional cohort studies with a pre-specified number of clinical variables, EHR is able to capture a broader range and larger number of confounding factors\citep{HouEtal2021ms}. In addition, multiple EHR features may be needed to represent one specific clinical variable, further amplifying the dimensionality of features need to capture candidate confounds. Modeling the propensity score (PS) along with the outcome regression (OR) in the presence of high dimensionality and potential model mis-specification is highly challenging. To the best of our knowledge, no methods currently exist to estimate ATE under the SSL setting when both the treatment group, denoted by $A$, and the response, denoted by $Y$, are only observed in a small subset of patients. For conciseness, we refer to this specific SSL setting as {\em double missing SSL}. In this paper, we fill in the methodology gap by proposing {\bf S}emi-supervised {\bf M}ultiple {\bf MA}chine {\bf L}earning  (SMMAL) estimators for ATE that leverage both the fully observed surrogates for $Y$ and $A$, denoted by $\bS$, and the partially observed gold standard labels on $Y$ and $A$.

Under the supervised setting where both $A$ and $Y$ are observed,
much progress have been made in recent years on ATE estimation
with confounding adjustment from machine-learning and/or high-dimensional regression.
In the low-dimensional setting, the estimation of ATE is a well studied problem including procedures that achieve  semi-parametric efficiency and double robustness \citep{RobinsRotnitzkyZhao94,BangRobins05}.
Extension to the high-dimensional setting, however, is not straightforward due to the slower convergence rates in the estimated model parameters and the difficulty imposed by not only the bias and variance trade-off in the process of regularization but also the inherent information theoretic barriers on obtaining fast enough estimation rates in high dimensional problems.   Similar challenges arise when incorporating more flexible machine-learning models to overcome model mis-specifications. Following intuitions parallel to the low-dimensional setting, flexible approaches for confounder adjustments have been proposed via modeling of PS and OR,
including $L_1$ regularized regression \citep{Farrell15}, neural network \citep{FarrellLM21}, and a general machine learning framework \citep{ChernozhukovEtal18DML}.
To overcome the high dimensionality, several methods require consistent estimation for PS and OR, which translates to proper model specification and sparsity for high-dimensional regressions \citep[e.g.]{BelloniCH14,LiuZZ21,HouBradicXu21ahazATE,BelloniCFH17}.
To improve the robustness,
\cite{Tan19AOSdr} designed a calibrated
estimation that leads to valid inference for the average treatment effect
even if one of the high-dimensional logistic PS or linear OR model
is mis-specified.
\cite{SmuclerEtal19} formalized the concept of double robustness in high-dimensional setting by defining the sparsity double robustness and model double robustness properties; and also generalized the idea of \cite{Tan19AOSdr} to a wide range of PS and OR models.
\cite{BradicEtal19} established a sharper sparsity double robustness property of the calibrated
estimation.
Unlike the double model approaches listed above, \cite{wang2020debiased} considered a single model approach in which they debiased the regularized PS model in the inverse probability of treatment weighting estimator to achieve $\sqrt{n}$-inference.

Semi-supervised estimation for ATE is less well studied with existing literature focusing almost entirely on the setting where $Y$ is observed for patients in the labeled set of size $n$ but $A$ and proxies of $Y$ along with confounders $\bX$ are observed for all subjects of size $N$, where $n \ll N$. Thus, the missingness proportion for $Y$ tends to 1 when $N \to \infty$, which sets it apart from standard analyses for classical missing data settings. For example, SSL estimators for the ATE have been proposed by
\cite{ChengACai21} when $Y$ is  missing-completely-at-random and by \cite{ZhangBradicChakrabortty2021} and \cite{KallusMao2020arxiv} when $Y$ is missing-at-random. However, these methods cannot be easily adapted to the setting where both $Y$ and $A$ are missing. The missingness in $A$ has a fundamental difference from the missingness in $Y$ since
treatment is an internal node in the causal pathway ``confounder-treatment-outcome''. Missing $A$ along with $Y$ triggers a complex reshuffle of the dependence structure.

In this paper, we propose an efficient and robust SSL estimator for ATE when both $Y$ and $A$ are only observed for the labeled subset but the confounders $\bX$ and surrogates for $Y$ and $A$ are observed for all $N$ patients. We derive the SMMAL estimator by first deriving the efficient influence function for the ATE under this SSL setting and then constructing a cross-fitted multiple machine learning  estimator. We subsequently provide a formal characterization of semi-parametric efficiency under the SSL setting with the SMMAL estimator coupled to B-spline regressions over low-dimensional space. We also design a doubly robust estimator with a two-layer cross-fitted calibrated estimation for high-dimensional logistic
PS and OR models.
Via cross-fitting and a truncation in calibration weights, we do not require the sparsity assumption in the first-layer estimation for PS and OR.
We further show that our doubly robust SMMAL estimator is adaptively doubly robust, rate doubly robust when both PS and OR models are correct and model doubly robust when one of them is correct. The SMMAL estimator also does not require correct specifications of the imputation models for $A$ and $Y$ given $\bW = (\bX\trans,\bS\trans)\trans$.

In view of the discussion above, we list our key contributions herein:
\begin{enumerate}
    \item We formalize the efficient estimation under SSL setting with
    a decaying labeling rate. Our theory justifies the efficiency claims of existing works and can provide benchmark for future work in this direction.  A discussion regarding the subtleties and challenges involved in this formalization and subsequent analyses can be found in Remark \ref{remark:SSLeff}.

    \item We lay out a general approach for efficient SSL
    with complex missing data structure. Our general approach is particularly convenient when the missingness-induced correlation renders typical projection approach difficult. A explanation of the challenge from missing treatment can be found in Remark \ref{remark:infl}, and the general framework is given in Section \ref{sect:eff}.
    \item We make progresses in the doubly robust estimation with high-dimensional confounders with an adaptive sparsity/model double robustness result and removal of sparsity requirement in initial estimation. The comparison with related work can be found in Remark \ref{remark:DR}.
\end{enumerate}

The remainder of the paper is organized as follows. We introduce our causal inference structure under the SSL setting along with the notations in Section \ref{sect:notation}.
In Section \ref{sect:method}, we first present the efficient influence function, followed by the multiple machine learning estimator and the model multiply robust estimator derived from the efficient influence function.
In Section \ref{sect:method}, we state the theoretical guarantees of the $\sqrt{n}$-inference on the average treatment effect from our methods, whose proofs are provided in the Supplementary Materials.
We also established the semi-parametric efficiency bound for the SSL of average treatment effect with missing treatment and outcome in low-dimensional space.
In Section \ref{sect:sim}, we assess the finite sample performance of our SSL methods and compare them to supervised benchmarks.
In Section \ref{sect:eff}, we offer the generalization of our work to the SSL
of other parameters under arbitrary missingness patterns.
Finally we conclude with a brief discussion in Section \ref{sect:discuss}.

\section{Setting and notation}\label{sect:notation}

For the $i$-th observation, $Y_i\in \R$ denotes the outcome variable,
$A_i \in \{0,1\}$ denotes the treatment group indicator,
$R_i \in \{0,1\}$ indicates whether $(Y_i,A_i)$ is observed,
$\bS_i\in \R^q$ denotes the surrogates for $Y_i$ and $A_i$,
and $\bX_i\in \R^{p+1}$ denotes the vector of (potential) confounders with the first element being $1$.
We use the notations without the subscript indices to denote the generic versions of these random variables. We observe $N$ independent and identically distributed (i.i.d.) observations, $\Dscr = \{ \bD_i = (R_i, R_iY_i, R_iA_i, \bW_i\trans)\trans, i = 1, ..., N\}$, where $\bW_i = (\bX_i\trans,\bS_i\trans)\trans$.
We assume that the labeled subjects are randomly sampled by design
\begin{equation}\label{aseq:MCAR}
  R \indep (Y, A, \bX, \bS)
\end{equation}
 and the number of labelled sample is $n$ with the proportion of labeled observation being $\rho = \E(R) \in (0,1)$ with $\rho \to 0$ as $N\to \infty$. We choose the missing-completely-at-random (MCAR) formulation \eqref{aseq:MCAR} to better connect with existing literature on semi-parametric estimation and missing data.   The results under MCAR, through concentration of measure arguments,  are largely applicable to the other formulation
 like first $n$ samples $R_i = \ind(i \le n)$ or sampling without replacement
 $$
 \P(R_1=r_1,\dots,R_N=r_N) = \prod_{i=1}^N \left(\frac{n-\sum_{j<i}r_j}{N-i+1}\right)^{r_i}
 \left(1-\frac{n-\sum_{j<i}r_j}{N-i+1}\right)^{1-r_i}, \,
 \sum_{i=1}^N r_i = n
 $$
 for a deterministic sequence $n$.

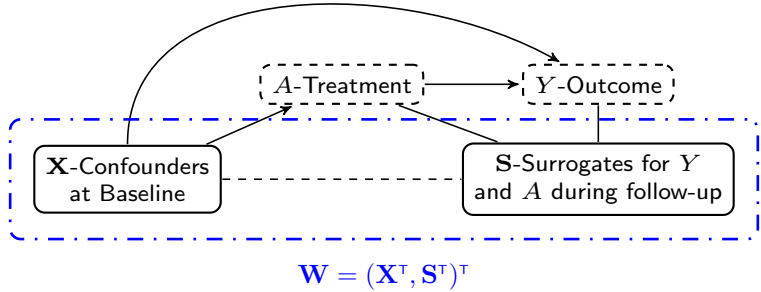
\begin{figure}
\centering
\begin{tikzpicture}[
  font=\sffamily\tiny,
  every matrix/.style={ampersand replacement=\&,column sep=0.5cm,row sep=0.5cm},
  observed/.style={draw,thick,rounded corners,fill=white,inner sep=.1cm,scale=1.5},
  unobserved/.style={draw,thick,rounded corners, dashed,fill=white,inner sep=.1cm,scale=1.5},
  sink/.style={source,fill=green!20},
  datastore/.style={draw,very thick,shape=datastore,inner sep=.3cm,scale=1.5},
  dots/.style={gray,scale=2},
  to/.style={->,>=stealth',shorten >=1pt,semithick,font=\sffamily\footnotesize},
  link/.style={semithick,font=\sffamily\footnotesize},
  every node/.style={align=center}]
  \matrix{
      \&  \node[unobserved] (A) {$A$-Treatment};
      \&  \node[unobserved] (Y) {$Y$-Outcome}; \\
   \node[observed] (X) {$\bX$-Confounders \\ at Baseline};
    \&
     \& \node[observed] (Sy) {$\bS$-Surrogates for $Y$ \\and $A$ during follow-up};\\
  };
  	\draw[to] (X) --  (A);
	\draw[to] (A) -- (Y);
  	\draw[to] (X) to [out=90,in=150](Y);
    \draw[link] (A) to (Sy); 
    \draw[link] (Y) to (Sy); 
    \draw[link, dashed] (X) to (Sy); 

  \tikzset{box dotted/.style={draw, line width=1pt,
                               dash pattern=on 1pt off 4pt on 6pt off 4pt,
                                inner sep=2mm, rectangle, rounded corners}};


    \node (W box) [box dotted,draw=blue, inner sep=3mm,
                            fit = (X) (Sy)] {};

    \node at (W box.south) [below, inner sep=3mm, text = blue,
                            font=\sffamily\footnotesize]
  { $\bW = (\bX^\top,\bS^\top)^\top$};

\end{tikzpicture}
\caption{Causal diagrams of SSL setting with missing treatment and outcome. The surrogates $\bS$ should associate with $A$ and $Y$ to provide information about them through the unlabeled data. The baseline confounders $\bX$ may or may not associate with $\bS$.
}\label{fig:dag}
\end{figure}

We define the ATE over the full population as
\begin{equation}\label{def:ate}
  \Delta_* = \E\left\{\E(Y \mid \bX, A=1) - \E(Y \mid \bX, A=0) \right\}.
\end{equation}
To ensure the proper causality of the definition in \eqref{def:ate},
we adopt the typical counterfactual outcome framework and its standard assumptions \citep{ImbensRubin15book,HernanRobins2023book}.
Let $Y(a)$ be the counterfactual outcome with treatment set as $a$, for $a \in \{0,1\}$
The ATE defined by the counterfactual outcome is
\begin{equation}\label{def:ate-Ya}
  \Delta_* = \E\left(Y^{(1)} - Y^{(0)}\right).
\end{equation}
Under the following assumptions, the definitions \eqref{def:ate} and \eqref{def:ate-Ya} are equivalent.
\begin{assumption}\label{assume:causal}
\begin{enumerate}[label = (\alph*)]
  \item Consistency: $Y = Y^{(A)}$;
  \item Ignorability: $\left(Y^{(1)},Y^{(0)}\right) \indep A \mid \bX$.
\end{enumerate}
\end{assumption}
The surrogates $\bS$ are not involved in the definition of the ATE,
as they are collected after treatment initiation.
We place no stringent causal assumption regarding $\bS$
and only require that $\bS$ is predictive for $A$ and $Y$.
We illustrate the causal relationship among the variables in Figure \ref{fig:dag}.
Throughout the paper, we assume \eqref{aseq:MCAR} and Assumption \ref{assume:causal}
without explicit reference.

\section{SMMAL Estimation}\label{sect:method}

We start by presenting in Section \ref{sect:sub-infl} the efficient influence function under the double missing SSL setting.
The  formalization of efficiency will be justified later
in Section \ref{sect:theory-eff}
and the term efficient influence function here has been heuristically borrowed from classical missing data setting
by existing literature in SSL setting \citep{ChengACai21,KallusMao2020arxiv}.
Then, we discuss the estimation of ATE with different ways of estimating the nuisance models involved
in the efficient influence function in Section \ref{sect:sub-np} for  low-dimensional $\bX$ and in Section \ref{sect:sub-highd} for high-dimensional $\bX$. Cross-fitting is a strategy adopted for both settings, where we split the data into $K$ (e.g. $K=10$) folds of approximately equal size.
For $k=1, ..., K$, we let $\Ical_k$ denote the index set for the $k$th fold of the data with size $N_k = |\Ical_k |$ and let $\Ical^c_k = \{1,\dots,N\}\setminus \Ical_k$, where $|\Ical|$ denotes the carnality of $\Ical$.

\subsection{The efficient influence function}\label{sect:sub-infl}

We define the following nuisance models:
\begin{alignat*}{3}
  &\text{PS: } &\quad& \P(A=a \mid \bX) = \pi(a,\bX), \,
   \quad  \text{OR: } &\quad & \E(Y \mid A =a, \bX) = \mu(a, \bX), \\
  &\text{Imputations: } &\quad& \P(A=a \mid \bW) = \Pi(a,\bW), \,
   &\quad & \E(Y \mid A=a, \bW) = m(a,\bW).
\end{alignat*}
We use the subscript star to indicate the true models, $\pi_*$, $\mu_*$, $\Pi_*$, $m_*$.
Starting from the efficient influence function with complete observation of treatment and outcome
\citep{RobinsRotnitzkyZhao94,KallusMao2020arxiv},
$$
\phi_{\cmp}(Y,A,\bX) = \mu_*(1,\bX) -  \mu_*(0,\bX)
+ \frac{\ind(A=1)}{\pi_*(1,\bX)}\{Y - \mu_*(1,\bX)\}
- \frac{\ind(A=0)}{\pi_*(0,\bX)}\{Y - \mu_*(0,\bX)\} - \Delta_*,
$$
we produce the efficient influence function through the following mapping
\begin{align}
 & \phi_{\SSL}(Y,A,\bW,R) \notag \\
 = & \E\{ \phi_{\cmp}(Y,A,\bX)\mid \bW\}
 + \frac{R}{\rho} [\phi_{\cmp}(Y,A,\bX) - \E\{ \phi_{\cmp}(Y,A,\bX)\mid \bW\}] \label{EIF1} \\
 = & \mu_*(1,\bX)
+ \frac{\Pi_*(1,\bW)}{\pi_*(1,\bX)}\{m_*(1,\bW) - \mu_*(1,\bX)\} \notag \\
&
-  \mu_*(0,\bX)
- \frac{\Pi_*(0,\bW)}{\pi_*(0,\bX)}\{m_*(0,\bW) - \mu_*(0,\bX)\} - \Delta_* \notag \\
& + \frac{R\{\ind(A=1)Y - \ind(A=1)\mu_*(1,\bX) - \Pi_*(1,\bW)m_*(1,\bW) + \Pi_*(1,\bW)\mu_*(1,\bX)\}}{\rho \pi_*(1,\bX)} \notag \\
& - \frac{R\{\ind(A=0)Y - \ind(A=0)\mu_*(0,\bX) - \Pi_*(0,\bW)m_*(0,\bW) + \Pi_*(0,\bW)\mu_*(0,\bX)\}}{\rho \pi_*(0,\bX)}. \label{def:infl}
\end{align}
In \eqref{EIF1}, $\E\{ \phi_{\cmp}(Y,A,\bX)\mid \bW\}$ is the maximal information on ATE from the unlabeled data with a known imputation model, and the second term is the price for training the best imputation model over the labeled data. We defer the rigorous derivation of this procedure to Section \ref{sect:eff}.

The efficient influence function in the missing data context is usually derived
by projecting an arbitrary initial influence function to the nuisance tangent space and subtract the projection from the initial influence function \citep{Tsiatis2007book}.
The approach has been applied to the SSL setting with missing outcome by first deriving the efficient influence function under missing data setting and then taking limit $n/N \to 0$ for SSL setting \citep{KallusMao2020arxiv}.
However, existing derivations of efficient influence function cannot be
directly applied to the SSL setting with $\rho$  tending to $0$ but not exactly $0$.
Moreover, the projection involved in deriving the efficient influence function becomes highly challenging if the missing component is a non-terminal node in the causal diagram.
In our setting, $\bS$ is the only terminal node, and both missing components $Y$ and $A$ are non-terminal nodes.
Additional challenges arise while conceptualizing semi-parametric efficiency in the $\rho\rightarrow 0$ setting. In the Section \ref{sect:theory-eff},  we address these issues by developing a asymptotic local minimax result similar to \cite{BegunEtal83} to demonstrate the optimality of our claimed efficient influence function. We further highlight the detailed technical challenges in deriving $\phi_{\SSL}$.

\subsection{SMMAL Procedure}\label{sect:sub-np}

Inspired by the double machine learning estimation \citep{ChernozhukovEtal18DML} based on $\phi_{\cmp}$,
we propose the following SMMAL estimator for ATE:
\begin{enumerate}
  \item
For each labelled fold $k$, we estimate the nuisance models by the out-of-fold data $\Ical_k^c$,
obtaining $\hat{\pi}\supk$, $\hat{\mu}\supk$, $\hat{\Pi}\supk$, $\hat{m}\supk$;

\item Estimate the ATE by
\begin{align}
   \hat{\Delta}_{\MML} = & \frac{1}{N}\sum_{k=1}^{K}\sum_{i \in \Ical_k} \hat{\Vcal}_{ik}, \notag \\
\hat{\Vcal}_{ik} = &    \hat{\mu}\supk(1,\bX_i)
+ \frac{\hat{\Pi}\supk(1,\bW_i)}{\hat{\pi}\supk(1,\bX_i)}\{\hat{m}\supk(1,\bW_i) - \hat{\mu}\supk(1,\bX_i)\} \notag \\
&
-  \hat{\mu}\supk(0,\bX_i)
- \frac{\hat{\Pi}\supk(0,\bW_i)}{\hat{\pi}\supk(0,\bX_i)}\{\hat{m}\supk(0,\bW_i) - \hat{\mu}\supk(0,\bX_i)\}  \notag \\
& + \frac{R_i\{A_iY_i - A_i\hat{\mu}\supk(1,\bX_i)\}}{\rho  \hat{\pi}\supk(1,\bX_i)}
- \frac{R_i\{(1-A_i)Y_i - (1-A_i)\hat{\mu}\supk(0,\bX_i) \}}{\rho  \hat{\pi}\supk(0,\bX_i)} \notag \\
&- \frac{R_i\{ \hat{\Pi}\supk(1,\bW_i)\hat{m}\supk(1,\bW_i) - \hat{\Pi}\supk(1,\bW_i)\hat{\mu}\supk(1,\bX_i)\}}{\rho  \hat{\pi}\supk(1,\bX_i)}
 \notag \\
& + \frac{R_i\{ \hat{\Pi}\supk(0,\bW_i)\hat{m}\supk(0,\bW_i) - \hat{\Pi}\supk(0,\bW_i)\hat{\mu}\supk(0,\bX_i)\}}{\rho  \hat{\pi}\supk(0,\bX_i)} . \label{def:ate-mml}
\end{align}
\item Estimate the variance of $\nhalf(\hat{\Delta}_{\MML}-\Delta_*)$ by
\begin{equation}
\hat{\Vcal}_{\MML}=\frac{\rho}{N}\sum_{k=1}^{K}\sum_{i \in \Ical_k} (\hat{\Vcal}_{ik}-\hat{\Delta}_{\MML})^2. \label{def:se-mml}
\end{equation}
\end{enumerate}
As $\rho\to 0$, $\hat{\Vcal}_{\MML}$ is dominated by the variability from terms with $R_i/\rho$.
The $(1-\alpha)\times 100\%$ confidence interval for ATE can be constructed with
$\hat{\Delta}_{\MML}$ and $\hat{\Vcal}_{\MML}$,
$$
\left[\hat{\Delta}_{\MML} -  \Zcal_{\sub \alpha/2}\sqrt{\hat{\Vcal}_{\MML}/n}, \,
\hat{\Delta}_{\MML} +  \Zcal_{\sub \alpha/2}\sqrt{\hat{\Vcal}_{\MML}/n} \right]
$$
where $\Zcal_{\sub \alpha/2}$ is the $1-\alpha/2$ quantile of standard normal distribution.

Similar to existing results in double machine learning literature, any estimators for the nuisance models with suitable rates of consistency can be used in our proposal as well.
For low-dimensional $\bW$ and smooth nuisance models, we can choose B-spline regression
with proper order and degrees.
Precise discussions on these rates, related conditions for general estimators
and relevant smoothness classes for B-spline regression are collected
 in Section \ref{sect:theory-eff}.

\subsection{Doubly Robust SMMAL Construction in high-dimensions}\label{sect:sub-highd}

We next discuss a specific construction of the SMMAL estimator when the dimensions $p$ and $q$ grow with $n$ and $p$ may be larger than $n$. We focus on the binary $Y$ and put the high-dimensional logistic regression models on the nuisance models
\begin{gather}
  \pi(1,\bX) = g(\bga\trans\bX); \; \mu(a,\bX) = g(\bgb_{\suba}\trans\bX), \, a=0,1; \notag \\
  \Pi(1,\bW) = g(\bgx\trans\bX); \; m(a,\bW) = g(\bgz_{\suba}\trans\bW), \, a=0,1, \notag
\end{gather}
 with $g(x) = 1/(1+e^{-x})$,  $\dg(x) = e^x/(1+e^{x})^2$, and $\ell(y,x) = \log(1+e^x)-yx$.
Other types of generalized linear models may also be derived accordingly.
Inspired by the model doubly robust estimation \citep{SmuclerEtal19},
we propose the following doubly robust SMMAL estimator for $\Delta_*$ under the above nuisance models, denoted by $\hat\Delta_{\DR}$:
\begin{enumerate}
  \item
For each labelled fold $k$, we estimate the imputation models by the Lasso over out-of-fold data $\Ical_k^c$,
\begin{alignat}{2}
 & \hat{\bgx}\supk = \argmin_{\bgx \in \R^{p+q+1}} \frac{\sum_{i\in \Ical_k^c}
 R_i \ell(A_i,\bgx\trans\bW_i)}{\sum_{i\in \Ical_k^c}  R_i} + \lambda_{\eta} \|\bgx\|_1, &\,& \lambda_{\eta}\asymp\sqrt{\log(p+q)/n}, \notag \\
  & \hat{\bgz}_{\suba}\supk = \argmin_{\bgz \in \R^{p+q+1}} \frac{\sum_{i\in \Ical_k^c}
  \ind(A_i=a)R_i\ell(Y_i,\bgz\trans\bW_i)}{\sum_{i\in \Ical_k^c} \ind(A_i=a) R_i} + \lambda_{\zeta} \|\bgz\|_1, &\,& \lambda_{\zeta}\asymp\sqrt{\log(p+q)/n}; \label{def:eta-zeta}
\end{alignat}
  \item
For each labelled fold pair $(k_1,k_2)$, we estimate the initial PS and OR models by the Lasso over out-of-two-folds data $\Ical_{k_1,k_2}^c = (\Ical_{k_1}\cup \Ical_{k_2})^c$,
{\small\begin{alignat}{2}
  &\hat{\bga}\supkk_{\init} = \argmin_{\bga \in \R^{p+1}} \frac{\sum_{i\in \Ical_{k_1,k_2}^c}
  R_i\ell(A_i,\bga\trans\bX_i)}{\sum_{i\in \Ical_{k_1,k_2}^c}  R_i} + \lambda_{\alpha,\init} \|\bga\|_1, &\; & \lambda_{\alpha,\init}\asymp\sqrt{\log(p)/n}, \notag \\
  & \hat{\bgb}_{\suba,\init}\supkk = \argmin_{\bgb \in \R^{p+1}} \frac{\sum_{i\in \Ical_{k_1,k_2}^c}
  \ind(A_i=a)R_i\ell(Y_i,\bgb\trans\bX_i)}{\sum_{i\in \Ical_{k_1,k_2}^c} \ind(A_i=a) R_i} + \lambda_{\beta,\suba,\init} \|\bgb\|_1, &\;& \lambda_{\beta,\suba,\init}\asymp\sqrt{\log(p)/n};
  \label{def:ab-init}
\end{alignat}}
\item For each labelled fold $k_1$, we calibrate the PS and OR models by cross-fitting within out-of-fold data $\Ical_{k_1}^c$,
{\footnotesize
\begin{gather}
  \hat{\bga}\supki_{\sub a} = \argmin_{\bga \in \R^{p+1}} \sum_{k_2\neq k_1}\sum_{i\in \Ical_{k_2}}\frac{R_i}{n}
  \dg\subtau\left(\bX_i\trans\hat{\bgb}_{\sub 1,\init}\supkk\right)\{(a-A_i) \bga^\top \bX_i + I(A_i=a)e^{(-1)^a\bga\trans\bX_i}\} + \lambda_{\alpha,a} \|\bga\|_1, \notag \\
   \hat{\bgb}_{\sub a}\supki = \argmin_{\bgb \in \R^{p+1}} \frac{\sum_{k_2\neq k_1}\sum_{i\in \Ical_{k_2}} \exp\subtau \left\{(-1)^a\bX_i\trans\hat{\bga}\supkk_{\init}\right\}
  I(A_i=a)R_i\ell(Y_i,\bgb\trans\bX_i)}{\sum_{i\in \Ical_{k_1}^c} I(A_i=a) R_i} + \lambda_{\beta,a} \|\bgb\|_1,
   \label{def:ab-cal}
\end{gather}}

\noindent where $\tau(x) = \sgn(x)(|x|\wedge 2M)$,
for any function $f(\cdot)$, we let $f\subtau(\cdot)$ denote $f\{\tau(x)\}$,
 $\lambda_{\alpha,a}$, $\lambda_{\beta,a}, \asymp\sqrt{\log(p)/n}$,.
\item Construct the nuisance model estimators:
\begin{gather}
  \hat{\pi}\supk(1,\bX_i) = g\subtau  ( \bX_i\trans  \hat{\bga}\supk_{\sub 1}), \,
  \hat{\pi}\supk(0,\bX_i) = g\subtau  ( -\bX_i\trans  \hat{\bga}\supk_{\sub 0}), \,
  \hat{\mu}\supk(a,\bX_i) = g  ( \bX_i\trans \hat{\bgb}\supk_{\suba}), \notag \\
  \hat{\Pi}\supk(a,\bW_i) =  g  ( \bW_i\trans \hat{\bgx}\supk), \,
  \hat{m}\supk(a,\bW_i) =  g ( \bW_i\trans \hat{\bgz}_{\suba}\supk);
  \label{def:MMR-modelest}
\end{gather}

\item Estimate the ATE by sending \eqref{def:MMR-modelest} to \eqref{def:ate-mml}, producing $\hat{\Delta}_{\DR}$.

\item Estimate the variance by sending \eqref{def:MMR-modelest} and $\hat{\Delta}_{\DR}$
to \eqref{def:se-mml}, producing $ \hat{\Vcal}_{\DR}$.
\end{enumerate}
The $(1-\alpha)\times 100\%$ confidence interval for ATE can be constructed with
$\hat{\Delta}_{\DR}$ and $\hat{\Vcal}_{\DR}$,
$$
\left[\hat{\Delta}_{\DR} -  \Zcal_{\sub \alpha/2}\sqrt{\hat{\Vcal}_{\DR}/n}, \,
\hat{\Delta}_{\DR} +  \Zcal_{\sub \alpha/2}\sqrt{\hat{\Vcal}_{\DR}/n} \right]
$$
where $\Zcal_{\sub \alpha/2}$ is the $1-\alpha/2$ quantile of standard normal distribution.
The doubly robust estimator $\hat{\Delta}_{\DR}$ can be viewed as a specific construction of
$\hat{\Delta}_{\MML}$ with specific estimators for the nuisance models.

The two-level cross-fitting in \eqref{def:ab-init} and  \eqref{def:ab-cal}, similar to those adopted in \citep{HouGuoCai2021}, has the advantage of having larger training set for each Lasso compared to the averaging after data splitting in \cite{SmuclerEtal19}.
If we choose $K=10$, we are able to use at least 80\% data while the data splitting in  \cite{SmuclerEtal19} may only use 45\% data.
Larger training sample allows the choice of smaller penalty factor thus reducing the bias.
Taking averaging after data splitting, however, cannot reduce bias.
The truncation $\tau$ at $M$ in \eqref{def:ab-cal} secures the overlapping property of
the initially estimated models.
Besides numerical stability, we can remove the sparsity condition associated with the initial estimator of the mis-specified model.

\section{Theoretical Properties of the SMMAL}

We establish the $\sqrt{n}$-consistency of $\hat{\Delta}_{\MML}$ and the honest asymptotic coverage of the confidence intervals with consistent estimation of PS and OR models in Section \ref{sect:theory-general}. In Section \ref{sect:theory-eff}, we derive  the asymptotic distribution of the SMMAL estimator $\hat{\Delta}_{\MML}$ and the subsequent matching lower bound to show its  semi-parametric efficiency in the low-dimensional $\bW$ case while using B-spline series estimators for nuisance regression models. For high-dimensional sub-Gaussian $\bX$ and $\bW$ and sparse nuisance models, we demonstrate in Section \ref{sect:theory-highd} that  $\hat{\Delta}_{\DR}$ is adaptive sparsity/model doubly robust with sparse nuisance models \citep{RotnitzkySmuclerRobins20Bka,SmuclerEtal19}.

\subsection{$\sqrt{n}$-inference}\label{sect:theory-general}

We require the following assumptions for nuisance models and the machine-learning estimators.
\begin{assumption}\label{assume:MML}
For a fixed constant $M$, we assume
\begin{enumerate}[label = (\alph*), ref = \ref{assume:MML}\alph*]
  \item \label{assume:MML-bdd} (Bounded response) almost surely $\sup_{i = 1,\dots,N}|Y_i|\le M$;
  \item \label{assume:MML-overlap} (Overlapping) almost surely $\sup_{i = 1,\dots,N}\sup_{a=0,1} 1/\pi_*(a,\bX_i) \le M$;
  \item \label{assume:MML-bddest} (Bounded estimators) almost surely $$\sup_{k = 1,\dots,K}\sup_{i \in \Ical_k}\sup_{a=0,1}\max\left\{ |1/\hat{\pi}\supk(a,\bX_i)|,  |\hat{\mu}\supk(a,\bX_i)|, |\hat{\Pi}\supk(a,\bW_i)|, |\hat{m}\supk(a,\bW_i)|\right\} \le M; $$
  \item \label{assume:MML-rate}(Rate of estimation)
  \begin{align*}
  \sup_{k=1,\dots,K}& \|\hat{\pi}\supk-\pi_*\|_2 + \|\hat{\mu}\supk-\mu_*\|_2 +
   \|\hat{\Pi}\supk-\bar{\Pi}\|_2 + \|\hat{m}\supk-\bar{m}\|_2 \notag \\
   & + \sqrt{n}\|\hat{\pi}\supk-\pi_*\|_2 \|\hat{\mu}\supk-\mu_*\|_2 = o_p(1)
  \end{align*}
  for some $\bar{\Pi}$ and $\bar{m}$ satisfying
  $\sup_{i = 1,\dots,N}\sup_{a=0,1}\max\left\{\bar{\Pi}(a,\bW_i), |\bar{m}(a,\bW_i)|\right\} \le M$, where for two models $h_1(a,\bW)$ and $h_2(a,\bW)$, we define 
\begin{equation}\label{def:MSE}
\|h_1 - h_2\|_2 = \max_{a\in\{0,1\}} \sqrt{\E[\{h_1(a,\bW) - h_2(a,\bW)\}^2]}.
\end{equation}
Here we use the $\ell_2$-norm notation because the MSE thus defined correspond to the $\ell_2$-estimation
error for model coefficients under parametric models.

  \item \label{assume:MML-var} (Stable variance)
  \begin{align*}
    \Vcal_* =  \Var&\left[\frac{AY - A\mu_*(1,\bX)}{\pi_*(1,\bX)}
- \frac{(1-A)Y - (1-A)\mu_*(0,\bX) }{\pi_*(0,\bX)}\right. \notag \\
&- \frac{\{ \bar{\Pi}(1,\bW)\bar{m}(1,\bW) - \bar{\Pi}(1,\bW)\mu_*(1,\bX)\}}{\pi_*(1,\bX)}
 \notag \\
& + \left.\frac{\{ \bar{\Pi}(0,\bW)\bar{m}(0,\bW) - \bar{\Pi}(0,\bW)\mu_*(0,\bW)\}}{\pi_*(0,\bX)}\right] \in [1/M, M].
  \end{align*}
\end{enumerate}
\end{assumption}
We establish the validity and asymptotic distribution of $\hat{\Delta}_{\MML}$ in the following theorem.
\begin{theorem}\label{thm:MML}
Under Assumption \ref{assume:MML}, 
$$
\sqrt{n/\hat{\Vcal}_{\MML}} (\hat{\Delta}_{\MML} - \Delta_*) \leadsto N(0,1),
$$
where ``$\leadsto$'' denotes convergence in distribution.
\end{theorem}
Assumption \ref{assume:MML-bdd} guarantees the boundedness of all nuisance models.
When $Y$ is binary, the models $\pi_*$, $\mu_*$, $\Pi_*$ and $m_*$ are all bounded by one.
Assumption \ref{assume:MML-overlap} is equivalent to the standard overlapping condition $\pi_*(1,\bX_i) \in [1/M, 1-1/M]$.
Assumption \ref{assume:MML-bddest} can be guaranteed by truncation of nuisance model estimators at $M$, which would not compromise the estimation accuracy under Assumptions \ref{assume:MML-bdd} and \ref{assume:MML-overlap}.
Assumption \ref{assume:MML-var} ensures the proper scaling of the asymptotic variance of $\hat{\Delta}_{\MML}$. As noted following \eqref{def:se-mml},
the term with the $R/\rho$ factor from labeled data in $\phi_{\SSL}$ dominates its variance
if $\rho \to 0$. The rate condition for the PS and OR models in Assumption \ref{assume:MML-rate} matches those for the double machine-learning estimator proposed in \citep{ChernozhukovEtal18DML}.
Based on that, we can allow the imputation models to be mis-specified and estimated at an arbitrary slow rate.

\subsection{Semi-parametric efficiency with low-dimensional confounder}\label{sect:theory-eff}

We next formally establish the semi-parametric efficiency lower bound under the
double missing SSL setting.
Consider the non-parametric model for $(\bW,RA,RY,R)$
\begin{align*}
  \Scal_{\SSL} = & \left\{ d\Pf(\bw,a,y,r) = \{\rho f(\bw,a,y)\}^r
  \{(1-\rho)f_{\bW}(\bw)\}^{(1-r)} d \nu_{\SSL}(\bw,a,y,r): \right. \notag \\
  & \quad f\text{ is density over }\Wcal \otimes \{0,1\}\otimes\Ycal,\left.\text{and }
  f_{\bW}(\bw) = \sum_{a\in\{0,1\}}\int_{y\in\Ycal} f(\bw,a,y) d\nu_y(y) \right\}
\end{align*}
for some measures $\nu_y$ over $\Ycal$, $\nu_w$ over $\Wcal$ and
$$
\nu_{\SSL}(\bw,a,y,r) = (\nu_w \times \delta_{\{0,1\}} \times \nu_y) (\bw,a,y)\times \delta_1(r) + \nu_w(w) \times \delta_0(r)
$$
where $\delta_{\Acal}$ is the counting/Dirac measure over the set $\Acal$.
Elements in $\Scal_{\SSL}$ can be indexed by the density $f$, and we denote the
true density as $f_*$ and the true model $\Pfs$.

\begin{remark}\label{remark:infl}
In existing work on ATE \citep{RobinsRotnitzkyZhao94,KallusMao2020arxiv},
the model $A\mid \bX$ provides no information on the $Y \mid A,\bX$, and thus would not be included in the nuisance tangent space.
In our setting, however, the surrogates induced a correlation between subspaces corresponding to $A\mid \bX$ and $Y \mid A,\bX$ in the nuisance tangent space,
which indicates that $A\mid \bX$ provides information on the $Y \mid A,\bX$ through the unlabelled data.
As the result, the geometry of the model tangent space is more complex, and the projection is no longer obvious.
See Section \ref{app-sub:score-space} of the Supplementary Materials for details.
\end{remark}

We denote the total variation norm as $\|\cdot\|_{\TV}$.
In the following theorem, we establish the semi-parametric efficiency lower bound for $\Delta$ under
$\Scal_{\SSL}$ in the form of a local minimax theorem obtained in the spirit of \cite{BegunEtal83}.
\begin{theorem}\label{thm:LB}
  Under Assumptions \ref{assume:MML-bdd}, \ref{assume:MML-bddest} and \ref{assume:MML-var},
  we have
  $$
 \liminf_{c \to \infty} \liminf_{N \to \infty}\inf_{\hat{\Delta}}
  \sup_{ \|f-f_*\|_{\TV} \le c/\sqrt{n}} \frac{\int n(\hat{\Delta} - \Delta_*)^2 d \prod_{i=1}^N \Pf (\bw_i,a_i,y_i, r_i)}{\rho\Var \{ \phi_{\SSL}(Y,A,\bW,R)\}}
  \ge 1.
  $$
\end{theorem}
\begin{remark}\label{remark:SSLeff}
Theorem \ref{thm:LB} offers one example that the semi-parametric efficiency bound (SEB) derived under the classical missing data setting can be generalized to the SSL setting with $\rho \to 0$. Previous attempts to formalize semi-parametric efficiency in a the SSL settings have assumed that the entire distribution of $\bW$ is known, i.e. $N = \infty$ and $\rho = 0$. Under the simplified SSL setting with $N=\infty$, the SEB can be derived by straightforward applications of standard results in classical semiparametric literature -- see e.g. \cite{van2000asymptotic}. Indeed, one more reason for considering this simplifying version is due to the possible ambiguity of defining regular estimators in high dimensional problems and thereby formalizing efficiency through the calibration of the best regular estimator. We bypass this conceptual difficulty by instead providing a result in the flavor of local asymptotic minimaxity  -- which operates on all possible estimators instead of restricting to the class of regular procedures.
\end{remark}

Taking the limit, the lower bound in Theorem \ref{thm:LB} approaches the asymptotic variance of the labeled data component in $\phi_{\SSL}$,
\begin{align*}
\lim_{\rho \to 0} \rho\Var \{ \phi_{\SSL}(Y,A,\bW,R)\} = & \Var\{\phi_{\cmp}(Y,A,\bX)-\E\{\phi_{\cmp}(Y,A,\bX) \mid \bW\}\} \\
= & \Var\{\phi_{\cmp}(Y,A,\bX)\} - \Var\{\E\{\phi_{\cmp}(Y,A,\bX) \mid \bW\}\}.
\end{align*}

From the representation above, we show that the efficiency gain from the unlabeled data with surrogates is given by the variance of the  $\phi_{\cmp}$ explained by the surrogates and confounders.
We have a general version of the theorem in Section \ref{sect:eff} which provides the rigorous theoretical ground for existing and future work on efficient SSL \cite[for example][Section 4]{KallusMao2020arxiv}.

The key idea of the proof is to construct the two-dimensional least favorable perturbation with different scale in two directions.
The first direction is proportional to $\phi_{\cmp}(Y,A,\bX) - \E\{\phi_{\cmp}(Y,A,\bX)\mid \bW\}$ and of size $\asymp 1/\sqrt{n}$,
which reflects the level of the information on $\Delta_*$ from the labels and should naturally scale with the number of expected labels.
The second direction is proportional to $\E\{\phi_{\cmp}(Y,A,\bX)\mid \bW\}$ and of size $\asymp 1/\sqrt{N}$,
which reflects the level of the information on $\Delta_*$ from the unlabelled data and should scale with the total sample size.
The design of the different scales ensures the tightness of  log-likelihood ratio between the perturbation and the truth, which would otherwise be degenerating or diverging.

We next show that the lower bound is attained under low-dimensional smoothness class models for the nuisance functions and can be operationalized by feeding B-spline regressions to $\hat{\Delta}_{\MML}$.
Suppose the confounders and surrogates are bounded continuous variables of fixed dimension, $\bW \in [-M,M]^{p+q}, \quad p+q < d \asymp 1$.
We measure the smoothness of the models by $\Hcal(f(\cdot))$ the H\"{o}lder class
defined in Definition \ref{def:Holder}, Section \ref{app:detail} of the Supplementary Materials.
\begin{assumption}\label{assume:Bspline}
For a fixed constant $M$, we assume
\begin{enumerate}[label = (\alph*), ref = \ref{assume:Bspline}\alph*]
\item \label{assume:Bspline-density} (Bounded density) the density functions for $\bX$ and $\bW$, $f_{\bX}(\bx)$ and $f_{\bW}(\bw)$, are bounded and bounded from zero,
    $$
    f_{\bX}(\bx) \in [1/M, M], \, \forall \bx \in [-M,M]^p, \,
    f_{\bW}(\bw) \in [1/M, M], \, \forall \bw \in [-M,M]^{p+q};
    $$
\item \label{assume:Bspline-smooth}
(Smooth models) the smoothness of the nuisance models observe
$$
\frac{1}{1+\Hcal(\pi_*(a,\cdot))/p} + \frac{1}{1+\Hcal(\mu_*(a,\cdot))/p} < 1, \,
\Hcal(\Pi_*(a,\cdot)) >0, \, \Hcal(m_*(a,\cdot)) >0, \, a=0,1.
$$
\end{enumerate}
\end{assumption}

\begin{corollary}\label{cor:UB}
Under Assumptions \ref{assume:MML-bdd}, \ref{assume:MML-overlap}, \ref{assume:MML-var} and \ref{assume:Bspline},
we may choose B-spline regressions with order
$$
\kappa \ge \max\{\Hcal(\pi_*(a,\cdot)), \Hcal(\mu_*(a,\cdot)), \Hcal(\Pi_*(a,\cdot)),
\Hcal(m_*(a,\cdot)):\, a=0,1\}-1,
$$
degrees
\begin{alignat*}{2}
&1/\pi(a,\cdot): \, n^{\frac{1}{1+\Hcal(\pi_*(a,\cdot))/p}}, &\;&
\mu(a,\cdot): \,  n^{\frac{1}{1+\Hcal(\mu_*(a,\cdot))/p}}, \\
&\Pi(a,\cdot): \, n^{\frac{1}{1+\Hcal(\Pi_*(a,\cdot))/(p+q)}}, &\;&
m(a,\cdot): \, n^{\frac{1}{1+\Hcal(m_*(a,\cdot))/(p+q)}}
\end{alignat*}
and truncation at $M$
for $\hat{\Delta}_{\MML}$ to achieve
$$
\sqrt{n} (\hat{\Delta}_{\MML} - \Delta_*)/\sqrt{\rho\Var \{ \phi_{\SSL}(Y,A,\bW,R)\}} \leadsto N\left(0,1\right).
$$
\end{corollary}
Since by Theorem \ref{thm:MML} $\hat{\Delta}_{\MML}$ attains the lower bound established in Theorem \ref{thm:LB},
it is indeed semi-parametric efficient.
At the same time, the lower bound in Theorem \ref{thm:LB} is the semi-parametric efficiency bound
for $\Delta_*$ under SSL setting $\Scal_{\SSL}$.

\subsection{Doubly robustness with high-dimensional confounder}\label{sect:theory-highd}

To describe the sparsity/model double robustness of $\hat{\Delta}_{\DR}$,
we define the asymptotic limits for Lasso estimators in \eqref{def:eta-zeta}-\eqref{def:ab-cal}.
\begin{gather}
  \bar{\bgx} = \argmin_{\bgx \in \R^{p+q+1}}
 \E\{ \ell(A_i,\bgx\trans\bW_i)\}, \;
   \bar{\bgz}_{\suba} = \argmin_{\bgz \in \R^{p+q+1}} \E\{
  \ind(A_i=a) \ell(Y_i,\bgz\trans\bW_i)\}, \notag \\
  \bar{\bga}_{\init} = \argmin_{\bga \in \R^{p+1}} \E\{\ell(A_i,\bga\trans\bX_i)\}, \;
   \bar{\bgb}_{\suba,\init} = \argmin_{\bgb \in \R^{p+1}} \E\{\ind(A_i=a)\ell(Y_i,\bgb\trans\bX_i)\}, \notag \\
  \bar{\bga}_a = \argmin_{\bga \in \R^{p+1}}\E\left[\dg\left(\bX_i\trans\bar{\bgb}_{\sub a,\init}\right)\{(a-A_i) \bga^\top \bX_i + I(A_i=a)e^{(-1)^a\bga\trans\bX_i}\}\right], \notag \\
   \bar{\bgb}_{\sub a} = \argmin_{\bgb \in \R^{p+1}} \E\left[ \exp\left\{(-1)^a\bX_i\trans\bar{\bga}_{\init}\right\}
  I(A_i=a)\ell(Y_i,\bgb\trans\bX_i)\right],
   \label{def:bar}
\end{gather}
We let $\|\cdot\|_0$ denote the sparsity of a vector and $\|\cdot\|_{\psi_2}$
denote the sub-Gaussian norm for random variables or vectors.
The detailed definition is given in Definition \ref{def:subG}, Section \ref{app:detail} of the Supplementary Materials.
\begin{assumption}\label{assume:MMR}
For constant $M$ independent of dimensions $n$, $N$, $p$, $q$,
\begin{enumerate}[label = (\alph*), ref = \ref{assume:MMR}\alph*]
  \item \label{assume:MMR-subG} (Sub-Gaussian and bounded covariates)
  the vector of confounders and surrogates is sub-Gaussian,
  $\sup_{\|\bv\|_2=1}\|\bv\trans\bW\|_{\psi_2} \le M$,
  and coordinate-wisely bounded $\|\bW\|_\infty \le M$ almost surely;
  \item \label{assume:MMR-inv} (Identifiability)
  the variance of $\bW$ is invertible
  $\inf_{\|\bv\|_2=1} \bv\trans \Var(\bW) \bv \ge 1/M$;
  \item \label{assume:MMR-overlap} (Overlapping)
  the true propensity scores and the asymptotic predictions of all models
  are bounded away from zero and one, almost surely,
  $$
  \pi_*(a,\bX)\in [1/M,1-1/M], \, \max\left\{|\bar{\bgx}\trans \bW|, |\bar{\bgz}_{\suba}\trans \bW|,
  |\bar{\bga}_{\suba}\trans \bX|, |\bar{\bgb}_{\suba}\trans \bX|: a=0,1\right\}
  \le M;
  $$
  \item \label{assume:MMR-dr} and one of the following:
  \begin{enumerate}[label =(\roman*), ref = \ref{assume:MMR-dr}-\roman*]
    \item \label{assume:MMR-ps} (PS correct)
    the propensity model is correct, $\E(A\mid \bX) = g(\bga_*\trans \bX)$
    and the dimensions satisfy
    \begin{align}\label{eq:rate-MMR-ps}
      & \frac{\left(\|\bar{\bgb}_{\sub 1}\|_0+ \|\bar{\bgb}_{\sub 0}\|_0\right) \log(p)+ \left(\|\bar{\bgx}\|_0 + \|\bar{\bgz}_{\sub 1}\|_0+ \|\bar{\bgz}_{\sub 0}\|_0\right) \log(p+q)}{n} \notag \\
      & +  \|\bga_*\|_0\left(\|\bga_*\|_0 + \|\bar{\bgb}_{\sub 1}\|_0+ \|\bar{\bgb}_{\sub 0}\|_0\right) \log(p)^2/n =o_p(1);
    \end{align}

    \item \label{assume:MMR-or} (OR correct) the OR model is correct, $\E(Y\mid A=a, \bX) = g(\bgb_{*,\suba}\trans \bX)$
    and the dimensions satisfy
    \begin{align}\label{eq:rate-MMR-or}
      & \frac{\left(\|\bar{\bga}_{\sub 1}\|_0+ \|\bar{\bga}_{\sub 0}\|_0\right) \log(p)+ \left(\|\bar{\bgx}\|_0 + \|\bar{\bgz}_{\sub 1}\|_0+ \|\bar{\bgz}_{\sub 0}\|_0\right) \log(p+q)}{n} \notag \\
      & +\sum_{a=0,1}\|\bgb_{*,\suba}\|_0\left(\|\bgb_{*,\suba}\|_0+ \|\bar{\bga}_{\suba}\|_0\right)\log(p)^2/n =o_p(1);
    \end{align}
     \item \label{assume:MMR-both} (both correct) both models are correct, $\E(A\mid \bX) = g(\bga_*\trans \bX)$ and $\E(Y\mid A=a, \bX) = g(\bgb_{*,\suba}\trans \bX)$
    and the dimensions satisfy
    \begin{align}\label{eq:rate-MMR-both}
      & \frac{\left(\|\bga_*\|_0 +\|\bgb_{*,\sub 1}\|_0+ \|\bgb_{*,\sub 0}\|_0\right) \log(p)+ \left(\|\bar{\bgx}\|_0 + \|\bar{\bgz}_{\sub 1}\|_0+ \|\bar{\bgz}_{\sub 0}\|_0\right) \log(p+q)}{n} \notag \\
      & +\|\bga_*\|_0\left(\|\bgb_{*,\sub 1}\|_0+ \|\bgb_{*,\sub 0}\|_0\right)\log(p)^2/n =o_p(1);
    \end{align}
  \end{enumerate}
\end{enumerate}
\end{assumption}

\begin{theorem}\label{thm:MMR}
  Under Assumption \ref{assume:MMR}, $\hat{\Delta}_{\DR}$ converges in distribution to a normal random
variable at $\sqrt{n}$-rate,
$$
\sqrt{n/\hat{\Vcal}_{\DR}} (\hat{\Delta}_{\DR} - \Delta_*) \leadsto N(0,1),
$$
where ``$\leadsto$'' denotes convergence in the distribution.
\end{theorem}

Besides the double robustness toward PS and OR, $\hat{\Delta}_{\DR}$ is also robust to the imputation models in addition to the double robustness for PS and OR.
Throughout Assumption \ref{assume:MMR-dr}, we require no model assumption on the imputation besides the sparsity condition for convergence in $L_2$-norm.

\begin{remark}\label{remark:DR}
Regarding the PS model and OR model,
our $\hat{\Delta}_{\DR}$ is both rate doubly robust \cite{RotnitzkySmuclerRobins20Bka} and model doubly robust \citep{SmuclerEtal19}.
When both models are correct,
the dimension condition \eqref{eq:rate-MMR-both} for the PS model and OR model in Assumption \ref{assume:MMR-both}
satisfies the condition for rate doubly robust, i.e. each sparsity obeying $\|\bga_*\|_0\ll n/\log(p), \|\bgb_{*,\suba}\|_0 \ll n/\log(p)$
and their product satisfying $\|\bga_*\|_0\|\bgb_{*,\suba}\|_0 \ll n/\log(p)^2$.
In the case of only one model is correct, our $\hat{\Delta}_{\DR}$ can still provide $\sqrt{n}$-inference, thus being model doubly robust.
By the truncation $\tau$ in \eqref{def:ab-cal},
we are able to completely remove the sparsity requirement of the mis-specified initial model
under the overlapping condition of Assumption \ref{assume:MMR-overlap}.
The general framework of \cite{SmuclerEtal19} would require all models in \eqref{def:bar}
being sparse.
\end{remark}

\section{Simulation}\label{sect:sim}

We have conducted extensive simulation studies to evaluate the finite sample performance of the
semi-supervised multiple machine learning
and multiply robust estimation methods.
Throughout the simulations, we set the total sample size $N=10000$, the number of labels $n=500$,
the number of repeats as 1000,
$q=2$ with one surrogate $S_A$ for $A$ and another $S_Y$ for $Y$.
We focus on the situation that $Y$ is also binary.
The surrogates for binary $A$ and $Y$ are generated from mixture Beta distribution:
\begin{gather*}
  S_A = A S_{A,1} + (1-A) S_{A,0}, \; S_Y = Y S_{Y,1} + (1-Y) S_{Y,0}, \\
  S_{A,1} \sim Beta(\alpha_A, 1), \, S_{A,0} \sim Beta(1,\alpha_A), \,
  S_{Y,1} \sim Beta(\alpha_Y, 1), \, S_{Y,0} \sim Beta(1,\alpha_Y).
\end{gather*}
The mixture Beta distribution mimics the outputs from phenotyping algorithms, which
often takes value between zero and one \citep{LiaoEtal2019MAP}.
We considered a list of values for $\alpha_A$ and $\alpha_Y$ (Table \ref{tab:sim-auc}), corresponding to different level of
prediction accuracy measured by area-under-curve (AUC) of the receiver operating characteristic (ROC).
Five values were considered for $\alpha_A$ and $\alpha_Y$, creating 25 two-way combinations
for each simulation setting.

\begin{table}
  \centering
  \begin{tabular}{c ccccc}
    \hline
\hline
Setting & OK & Reasonable & Good & Great & Perfect \\
\hline
AUC & 0.80 & 0.90 & 0.95 & 0.99 & 0.999 \\
$\alpha$ & 1.84 & 2.39 & 2.99 & 4.26 & 5.88 \\
\hline
    \hline
  \end{tabular}
  \caption{List of parameters used in the mixture Beta distribution for the surrogates.}\label{tab:sim-auc}
\end{table}

We consider two scenarios for generating the data,
the low-dimensional smooth model and high-dimensional logistic regression.

\begin{figure}
\begin{center}
\includegraphics[width=0.4\textwidth]{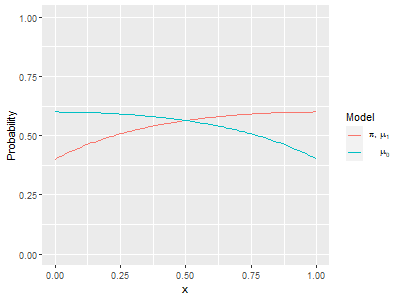}
\includegraphics[width=0.4\textwidth]{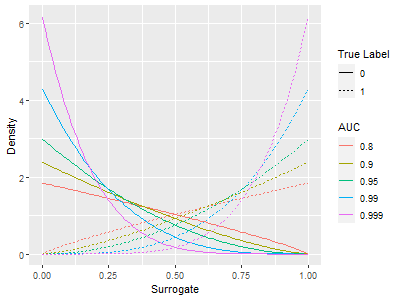}
\end{center}
\caption{Visualized simulation settings. Left-the models for PS and OR under the low-dimensional setting. Right-the mixture Beta distribution for surrogates at
different level of prediction accuracy (AUC 0.8, 0.9, 0.95, 0.99, 0.999).}\label{fig:sim-setting}
\end{figure}

\paragraph{Low-dimensional smooth model}
We generate the one dimensional $X \in \R$ from Uniform(0,1)
and set the PS and OR to be the following smooth models (Figure \ref{fig:sim-setting}):
$$
\pi_*(1,\bX) = \mu_*(1,\bX) = 1-1.2/(3-x^2), \; \mu_*(0,\bX) = 1-1.2/\{3-(1-x)^2\}.
$$
We used tensor product first order B-spline (piece-wise linear splines) regression
to estimate the nuisance models.
The splines were constructed from \emph{bs} function of the \emph{splines} R package.
The degrees were selected by 10 fold cross-validation among
integers less than $\sqrt{n} \approx 22$ according to the out-of-fold
entropy.
Using the cross-fitted nuisance models from B-spline regression with $K=10$,
we obtained point and interval estimates for the ATE based on $\hat{\Delta}_{\MML}$
and $\hat{\Vcal}_{\MML}$.
As the benchmark,
we also estimated the ATE using the labeled data only
by the double machine learning method \citep{ChernozhukovEtal18DML}.

\paragraph{High-dimensional logistic regression}
We generate the high-dimensional $\bX \in \R^{500}$ with $p=500$
from the multivariate Gaussian distribution with auto-regressive correlation structure:
$$
U_1,\dots, U_p \stackrel{i.i.d.}{\sim} N(0,1), \,
X_1 = U_1, \,
X_{j} = 0.5 X_{j-1} + \sqrt{0.75} U_j.
$$
We generate $A$ and $Y$ from the high-dimensional logistic regression models
\begin{align*}
  \text{Correct PS: } &
  \pi_*(1,\bX) = g(0.5 X_1 + 0.25 X_2 + 0.125 X_3); \\
\text{Wrong PS: }&
  \pi_*(1,\bX) = g\{(0.5 X_1 + 0.25 X_2 + 0.125 X_3)
  ( 1 + 0.0625 X_1 + 0.125 X_2 - 0.5 X_3)\}; \\
  \text{Correct OR: } &
  \mu_*(1,\bX) = g(0.1+0.25 X_1 + 0.125 X_2 + 0.0625 X_3),\\
  & \mu_*(0,\bX) = g(-0.1-0.25 X_1 - 0.125 X_2 - 0.0625 X_3); \\
  \text{Wrong OR: } &
  \mu_*(1,\bX) = g\{(0.1+0.25 X_1 + 0.125 X_2 + 0.0625 X_3) \\
   & \qquad\qquad\qquad \times ( 1 + 0.0625 X_1 + 0.125 X_2 - 0.5 X_3)\},\\
  & \mu_*(0,\bX) = g\{(-0.1-0.25 X_1 - 0.125 X_2 - 0.0625 X_3)\\
  & \qquad\qquad\qquad \times ( 1 + 0.0625 X_1 + 0.125 X_2 - 0.5 X_3)\}.
\end{align*}
The wrong models contain a second order interaction.
We consider 3 combinations corresponding to the three settings of Assumption \ref{assume:MMR-dr}:
correct models (correct PS + correct OR);
mis-specified PS (wrong PS + correct OR);
mis-specified OR (correct PS + wrong OR).
We set the number of the folds as 10
and fitted the imputations \eqref{def:eta-zeta} and initial estimators \eqref{def:ab-init}
using \emph{glmnet} from R-package \emph{glmnet}.
We fitted the calibrated estimators  \eqref{def:ab-cal} using \emph{rcal} from from R-package \emph{rcal}.
The penalty parameters were selected by 10-fold cross validation with out-of-fold entropy.
Using the cross-fitted nuisance models,
we estimated the ATE using $\hat{\Delta}_{\DR}$
and construct the 95\% confidence interval based on the variance estimator  $\hat{\Vcal}_{\DR}$.
As the benchmark,
we also estimated the ATE using the labeled data alone
by the model doubly robust estimation \citep{SmuclerEtal19}.

\begin{figure}
\begin{center}
  \includegraphics[width=0.8\textwidth]{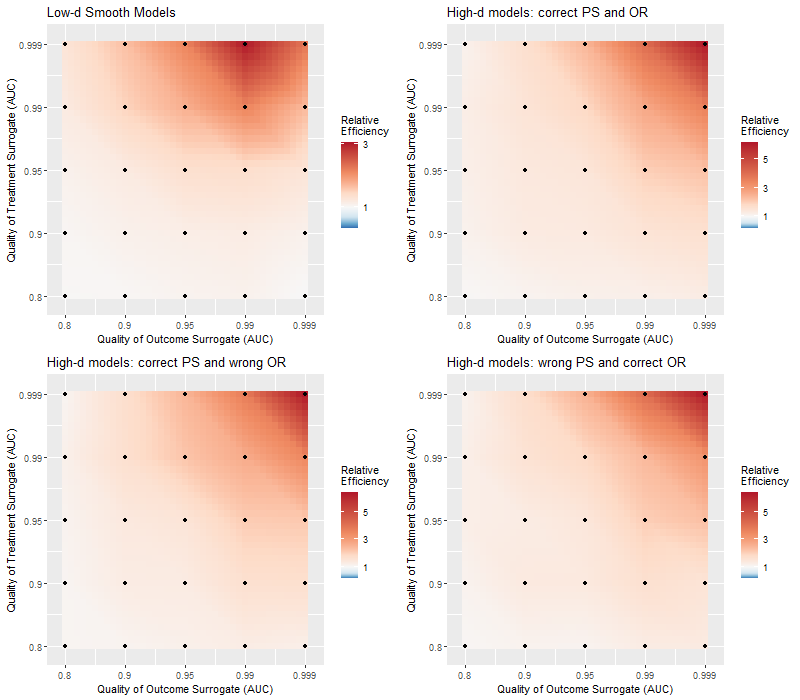}
\end{center}
\caption{Heat map for relative efficiency of the semi-supervised estimation compared to the benchmark supervised estimation in all four simulation settings. Deeper red indicates larger advantage of the semi-supervised estimation. We set relative efficiency one as white in all plots, but the scale varies between low-dimensional setting and high-dimensional settings. }\label{fig:sim-RE}
\end{figure}

\paragraph{Results}
In Figure \ref{fig:sim-RE}, we visualize the relative efficiency of our semi-supervised $\hat{\Delta}_{\MML}$, $\hat{\Delta}_{\DR}$ compared to their supervised benchmarks.
In general, our semi-supervised approaches gain efficiency from the unlabeled data
whose magnitude is increasing with the minimal prediction accuracy of the two surrogates.
With good imputation (AUC .95) from both surrogates, the relative efficiency is about 1.36-1.50 across
all settings.
With great imputation (AUC .99) from both surrogates, the relative efficiency is about 2.15-2.71 across
all settings.
The detailed simulation results containing the bias, standard deviation, average standard error,
coverage of 95\% confidence interval for our semi-supervised $\hat{\Delta}_{\MML}$,  $\hat{\Delta}_{\DR}$ along with those for the supervised benchmarks were presented in Tables
\ref{tab:sim-np}-\ref{tab:sim-hd-misOR} in Section \ref{app:sim} of the Supplementary Materials.
Our semi-supervised $\hat{\Delta}_{\MML}$,  $\hat{\Delta}_{\DR}$ achieved reasonably honest inference with coverage of 95\% confidence interval close to the nominal level.

\section{Semi-parametric efficient SSL}\label{sect:eff}

Our methodology developed for SSL of ATE with missing
treatment and outcome can be generalized for other SSL of other parameters with different missing data.
Consider the model for data $(R, R\bZ,\bW)$ with always observed $\bW$ and missing completely at random $\bZ$,
\begin{align*}
  \Scal_{\SSL} = &\left\{d\Pf(r,\bz,\bw,r) = [\rho f(\bz,\bw)]^r \left[(1-\rho)\int_{\bz\in\Zcal} f(\bz,\bw) d \nu_z(\bz)\right]^{(1-r)}
  d\nu_{\SSL}(r,\bz,\bw)
  :  \right. \\
  &\qquad \left. f(\bz,\bw)d\nu_{\cmp}(\bz,\bw) \in \Scal_{\cmp}\right\}
\end{align*}
for some complete data model class $\Scal_{\cmp}$ over $\Zcal \otimes \Wcal$ and measures $\nu_z$ over $\Zcal$, $\nu_w$ over $\Wcal$
and
$$
\nu_{\cmp} = \nu_z \times \nu_w, \;
\nu_{\SSL}(r,\bz,\bw) = \delta_1(r)\times \nu_{\cmp}(\bz,\bw)
+ \delta_0(r)\times\nu_w(\bw)  .
$$
Let $\Hscr$ is nuisance tangent space of $\Scal_{\SSL}$ at the true model $d\Pfs$ with $f=f_*$.
Suppose $\bgps_{\cmp}(\bZ,\bW)$ is the efficient influence function for parameter $\bgth$
under complete data model $\Scal_{\cmp}$.
\begin{assumption}\label{assume:eff}
For absolute constant $M$,
\begin{enumerate}[label = (\alph*), ref = \ref{assume:eff}\alph*]
  \item \label{assume:eff-MCAR} (Missing completely at random) $R \indep (\bZ,\bW)$;
  \item \label{assume:eff-label_info} (Informative labels) $\inf_{\|\bv\|_2=1}
  \bv\trans\Var\left[ \bgps_{\cmp}(\bZ, \bW) - \E\{\bgps_{\cmp}(\bZ, \bW) \mid \bW\}  \right]\bv \ge 1/M$;
 \item \label{assume:eff-W_flexible} (Model flexibility) $\E_*\{\bgps_{\cmp}(\bZ, \bW) \mid \bW\} \in \Hscr$;
  \item \label{assume:eff-bound_infl} (Boundedness of influence function)
  $\|\bgps_{\cmp}(\bZ, \bW)\|_2 \le M$ almost surely.
\end{enumerate}
\end{assumption}
We may derive the SSL efficient influence function by the following proposition.
\begin{proposition}\label{prop:SSL-eff}
Under Assumptions \ref{assume:eff-MCAR} and \ref{assume:eff-W_flexible},
the efficient influence function under SSL model $\Scal_{\sf SSL}$
is
\begin{equation}
  \bgps_{\SSL}(R, \bZ, \bW) = \frac{R}{\rho} \left[\bgps_{\cmp}(\bZ, \bW) - \E\{\bgps_{\cmp}(\bZ, \bW)\mid \bW\} \right] +  \E\{\bgps_{\cmp}(\bZ, \bW)\mid \bW\}.
\end{equation}
\end{proposition}
The influence function $\bgps_{\SSL}$ leads to a semi-parametric efficiency lower bound.
\begin{theorem}\label{thm:minimax}
  Under Assumptions \ref{assume:eff-MCAR}-\ref{assume:eff-bound_infl},
  we have the minimax semi-parametric efficiency for SSL of $\bgth$
  under $\Scal_{\sf SSL}$,
  $$
 \liminf_{c \to \infty} \liminf_{N \to \infty}\inf_{\hat{\bgth}}
  \sup_{ \|f-f_*\|_{\TV} \le c/\sqrt{n}} \frac{\int n\{\ba^\top(\hat{\bgth} - \bgth^*)\}^2 d \prod_{i=1}^N \P_{f} (\bz_i,\bw_i,r_i)}{\ba^\top\Var \{\rho \bgps_{\SSL}(R,\bZ,\bW)\}\ba}
  \ge 1.
 $$
\end{theorem}
Upper bound depends on the context. Like Corollary \ref{cor:UB},
the bound can be attained if nonparametric estimation of nuisance models
admit sufficiently fast rate of consistency,
which has been thoroughly studied under classical low-dimensional settings by \cite{Stone77,Stone82}.
While we focus on $\rho \to 0$ and $n \ll N$ setting, the theory also applies to classical setting with $\rho \in [1/M, 1-1/M]$ and $n \asymp N$ setting.

\section{Discussion}\label{sect:discuss}

Motivated by the increasing interest of generating real-world evidence for treatment effect with big yet noisy EHR data, we propose a robust and efficient
semi-supervised estimator for ATE under the doubly missing SSL setting. The SMMAL estimator gains efficiency by leveraging the large unlabelled data containing noisy yet predictive surrogates for $Y$ and $A$ with almost no additional requirement
than those needed for the supervised analysis using the labeled set alone. We established semi-parametric efficiency bound for the ATE estimator under the low dimensional confounder setting and constructed a doubly robust SMMAL estimator for the high dimensional confounder setting.

Our doubly robust estimation can be generalized to other models
if the calibrated estimation for the model is available.
For example, we can directly adopt the estimators from \cite{Tan19AOSdr} for linear outcome model.
The calibrated estimation is, however, limited to M-estimator in high-dimensional regression
due to the paucity of works on Z-estimators in high-dimensional setting.
It would be interesting to study if the Z-estimator approach \citep{VV15} can
be generalized to high-dimensional setting.

\bibliographystyle{chicago}
\bibliography{double_surrogate}

%
%
%
%
\end{document}